# Equation of state in the generalized density scaling regime studied from ambient to ultra-high pressure conditions

A. Grzybowski,* K. Koperwas, and M. Paluch

*Institute of Physics, University of Silesia, Uniwersytecka 4, 40-007 Katowice, Poland*

In this paper, based on the effective intermolecular potential with well separated density and configuration contributions and the definition of the isothermal bulk modulus, we derive two similar equations of state dedicated to describe volumetric data of supercooled liquids studied in the extremely wide pressure range related to the extremely wide density range. Both the equations comply with the generalized density scaling law of molecular dynamics versus $h(\rho)/T$ at different densities $\rho$ and temperatures $T$, where the scaling exponent can be in general only a density function $\gamma(\rho) = d\ln h / d\ln \rho$ as recently argued by the theory of isomorphs. We successfully verify these equations of state by using data obtained from molecular dynamics simulations of the Kob-Andersen binary Lennard-Jones liquid. As a very important result, we find that the one-parameter density function $h(\rho)$ analytically formulated in the case of this prototypical model of supercooled liquid, which implies the one-parameter density function $\gamma(\rho)$, is able to scale the structural relaxation times with the value of this function parameter determined by fitting the volumetric simulation data to the equations of state. We also show that these equations of state properly describe the pressure dependences of the isothermal bulk modulus and the configurational isothermal bulk modulus in the extremely wide pressure range investigated by the computer simulations. Moreover, we discuss the possible forms of the density functions $h(\rho)$ and $\gamma(\rho)$ for real glass formers, which are suggested to be different from those valid for the model of supercooled liquid based on the Lennard-Jones intermolecular potential.

*\* Corresponding author's email: andrzej.grzybowski@us.edu.pl*

## I. INTRODUCTION

In the last decade, our understanding of the glass transition and related phenomena has been considerably enhanced by making the observation that values of dynamic quantities such as structural relaxation time τ and viscosity η collected for many systems near the glass transition in various conditions of temperature $T$ and density $\rho(T,p)$ at ambient and elevated pressure $p$ can be plotted on one master curve according to a power law density scaling function $f(\rho^\gamma/T)$ with only one parameter γ, which is characteristic of a given material independent of thermodynamic conditions.[1,2,3,4,5,6,7,8,9,10,11,12,13,14,15] This scaling rule established phenomenologically has been also subjected to intensive theoretical and simulation studies,[16,17,18,19,20,21,22,23] which have led to a rather general conclusion that an effective short-range potential, $U_{eff}(r) = U_{IPL}(r) - A_t$ (where its repulsive part is given by an inverse power law (IPL) term, $U_{IPL}(r) = 4\varepsilon(\sigma/r)^{m_{IPL}}$ with $m_{IPL} \approx 3\gamma$, and $A_t$ is some small constant or linear attractive background), underlies the $\rho^\gamma/T$-scaling of the molecular dynamics near the glass transition. As a reference for the systems that complies with the power law density scaling rule at least to a good approximation, Dyre's group introduced a concept of *strongly correlating liquids*, which are defined by a strong linear correlation between isochoric equilibrium fluctuations of virial $W$ and potential energy $U$ with the correlation slope that corresponds to the scaling exponent γ, assuming that the $WU$ correlation is strong if the correlation coefficient $R > 0.9$.[18,19,20]

The simple and tempting theoretical grounds for the power law density scaling caused that this pattern of scaling has become a prominent trend in the investigations of glass formers. Nevertheless, already in the early reports on this matter,[4,17,24,25] a general form of the density scaling function has been considered, according to which, for instance, the structural relaxation time can obey the following scaling equation

$$\tau(\rho,T) = f\left(\frac{h(\rho)}{T}\right) \tag{1}$$

where the only density dependent function $h(\rho)$ is not limited to the power function $\rho^\gamma$. Thorough analyses of molecular dynamics (MD) simulation data confirmed[19,22,23,26] that the scaling exponent $\gamma$ is not constant and can depend at least on density even in simple models based on the Lennard-Jones (LJ) potential such as the Kob-Andersen binary Lennard-Jones (KABLJ) mixtures.[27] It has been realized that the pure power law density scaling is valid only for potentials with a single IPL term, however, a hidden scale invariance of molecular dynamics characterized by more complex potentials can be revealed by using reduced units[23,28,29,30] (e.g. the structural relaxation time in the reduced units, $\tilde{\tau} = \tau\rho^{1/3}T^{1/2}$, for τ, ρ, T in the usual LJ units). For instance, in the KABLJ liquid which is a double IPL model, the reduced units enable to scale the structural relaxation times versus $\rho^\gamma/T$ with $\gamma = const$ but only in the sufficiently narrow density range (e.g., the $\rho^\gamma/T$-scaling of $\tilde{\tau}$ in the KABLJ model occurs if the particle number density ρ ($\equiv N/V$, where $N$ is the particle number and $V$ is the system volume) ranges from 1.2 to 1.6, but it is not possible if ρ varies from 1.2 to 2.0).[31]

The general pattern of the density scaling (Eq. (1)) can be better understood within the theory of isomorphs recently formulated by Dyre's group.[29,32,33] According to this theory (Appendix A in Ref. 29), a system is *strongly* *correlating* if and only if it has isomorphs to a good approximation in its phase diagram, which are curves of isomorphic state points in the following sense: two state points $(T_1, \rho_1)$ and $(T_2, \rho_2)$ are isomorphic if all pairs of their physically relevant microconfigurations $(\mathbf{r}_1^{(1)}, \ldots, \mathbf{r}_N^{(1)})$ and $(\mathbf{r}_1^{(2)}, \ldots, \mathbf{r}_N^{(2)})$ characterized by identical reduced coordinates $\tilde{\mathbf{r}}_i^{(1)} = \tilde{\mathbf{r}}_i^{(2)}$ (where $\tilde{\mathbf{r}}_i = \rho^{1/3} \mathbf{r}_i$) have proportional configurational NVT Boltzmann factors, $\exp[-U(\mathbf{r}_1^{(1)}, \ldots, \mathbf{r}_N^{(1)})/k_B T_1] = C_{12} \exp[-U(\mathbf{r}_1^{(2)}, \ldots, \mathbf{r}_N^{(2)})/k_B T_2]$, where the constant $C_{12}$ depends only on the state points $(T_1, \rho_1)$ and $(T_2, \rho_2)$, not on the microscopic configurations. An undoubted achievement of this theory is an incontrovertible evidence[34] for the only density dependent scaling exponent $\gamma(\rho)$, which can be derived in the case of *strongly correlating systems* from the function $h(\rho)$ in Eq. (1) by logarithmic differentiating with respect to $\ln \rho$

$$\gamma = \frac{d \ln h}{d \ln \rho} \qquad (2)$$

Thus, the general density scaling idea given by Eq. (1) can be considered as a consequence of the generalization about the power density function $h(\rho) = \rho^\gamma$ with $\gamma = const$, which assumes that the scaling exponent γ can depend on density ρ based on Eq. (2). It should be noted that the scaling exponent argued to be dependent on density well corresponds to the density dependent slope of the WU correlation established from MD simulations in the KABLJ model. This result of simulation experiments can be well grounded and generalized within the theory of isomorphs, which enables to prove[34] that *strongly correlating systems* in general obey the configurational Grüneisen equation of state[35,36,37]

$$W = \gamma(\rho) U + \Phi(\rho) \qquad (3)$$

Although a general discussion on thermodynamics of condensed matter with strong pressure-energy correlations has been already done, involving Eq. (3)[34] and its version[32] for the double IPL potential model, until recently, no equation of state (EOS) has been formulated in the form convenient to describe PVT data of the *strongly correlating systems* considered in the general density scaling case characterized by the density scaling criterion[17,31,34]

$$\frac{h(\rho)}{T} = const \qquad (4)$$

which should be validated at $\tilde{\tau} = const$ if we analyze the density scaling of structural relaxation times in terms of Eq. (1) where τ should be also replaced with $\tilde{\tau}$.[31,34]

In this paper, we derive an EOS, which can be considered as an approximate PVT representation of the configurational Grüneisen EOS (Eq. (3)) for systems that meet the general density scaling criterion (Eq. (4)). We test a version of the EOS for the LJ potential using PVT data from our MD simulations in the KABLJ model and discuss a possible form of the EOS for real glass formers measured in the extremely wide pressure range that corresponds to the wide density range, in which the density scaling law given by Eq. (1) with the density function $h$ approximated by $h(\rho) = \rho^\gamma$ with $\gamma = const$ is not sufficient to scale molecular dynamics of the materials, but it is possible to find another density dependent scaling function $h(\rho)$ that enables the scaling in terms of Eq. (1) and implies the density dependent scaling exponent $\gamma(\rho)$ based on Eq. (2).

## II. EQUATION OF STATE FOR PVT DATA OF *STRONGLY CORRELATING LIQUIDS*

A few years ago, we suggested an equation of state derived in the power law density scaling regime first in its isothermal form,[38,39] which has been later generalized[40] to describe PVT in a convenient way not limited only to any constant temperature condition,

$$p^{conf} = p_0^{conf} + \frac{B_T^{conf}(p_0^{conf})}{\gamma_{EOS}}\left[\left(\frac{\rho}{\rho_0}\right)^{\gamma_{EOS}} - 1\right] \tag{5}$$

where $\rho_0^{-1} = \upsilon(T, p_0^{conf}) = A_0 + A_1(T-T_0) + A_2(T-T_0)^2$ and the configurational isothermal bulk modulus $B_T^{conf}(p_0^{conf}) = B_{T_0}^{conf}(p_0^{conf})\exp(-b_2^{conf}(T-T_0))$ at a reference configurational pressure, $p_0^{conf} = p^{conf}(T_0, p_0) = p_0 - RT_0\rho(T_0, p_0)/M$, which is established at the reference temperature $T_0$ and pressure $p_0$ using the system molar mass $M$ and the gas constant $R$. We determined[26,38,39,40] the physical meaning of the exponent $\gamma_{EOS}$ by arguing the relation $\gamma_{EOS} \cong m_{IPL}/3$, where $m_{IPL}$ is the exponent of the repulsive IPL term of the effective short-range intermolecular potential, $U_{eff}(r) = 4\varepsilon(\sigma/r)^{m_{IPL}} - A_t$, which is suggested to be responsible for the power law density scaling of molecular dynamics near the glass transition.

Based on the definition of the isothermal bulk modulus, $B_T = (\partial \ln \rho/\partial p)_T$, we also arrived at another EOS, finding its isothermal[39] and generalized[40] versions which comply with the linear pressure dependence, $B_T(p) = B_T(p_0) + \gamma_{EOS}(p-p_0)$, where $p_0$ is a reference pressure. We showed that the fitting of PVT data to this EOS

$$p = p_0 + \frac{B_T(p_0)}{\gamma_{EOS}}\left[\left(\frac{\rho}{\rho_0}\right)^{\gamma_{EOS}} - 1\right] \qquad (6)$$

where $\rho_0^{-1} = v(T, p_0) = A_0 + A_1(T-T_0) + A_2(T-T_0)^2$ and the isothermal bulk modulus $B_T(p_0) = B_{T_0}(p_0)\exp(-b_2(T-T_0))$ at a reference pressure $p_0$, yields a similar value of the exponent $\gamma_{EOS}$ to that found by using Eq. (5) in case of a tested system. Such results have been obtained for both the simulation and experimental data.[26,39,40] Moreover, our analyses of the PVT data collected from MD simulations in the KABLJ model and its version limited to the repulsive IPL term confirmed the relation $\gamma_{EOS} \cong m_{IPL}/3$ for both Eqs. (5) and (6) in the case of the simple models based on the Lennard-Jones intermolecular potential.[26] It is worth noting that very recently we have also derived[41] a similar equation of state for the activation volume in an analogous way to that employed in the derivation of Eq. (6), but using the definition of the isothermal bulk modulus for the activation volume.

The already mentioned recent investigations[31,34] of the generalized density scaling within the framework of the theory of isomorphs have induced us to find the counterparts of Eqs. (5) and (6) not limited to the power law density scaling. We have pointed out that one can do that by an appropriate generalization about the method exploited[38,42] to derive Eq. (5) from the effective short-range intermolecular potential, $U_{eff}(\mathbf{R}) = U_{IPL}(\mathbf{R}) - A_t$, which involves all particle coordinates denoted by $\mathbf{R}$ at a given state $(T,\rho)$ and implies a simple relation, $U_{eff}\left((\rho/\rho_0)^{-1/3}\mathbf{R}\right) = (\rho/\rho_0)^{m_{IPL}/3} U_{IPL}(\mathbf{R}) - A_t$, based on the Euler theorem on homogeneous functions. One can note[43] that the generalized density scaling requires the following modification of this relation

$$U_{eff}\left((\rho/\rho_0)^{-1/3}\mathbf{R}\right) = h(\rho)U(\mathbf{R}) + g(\rho) \qquad (7)$$

which is assumed to well separate density and configuration contributions ($h(\rho)$, $g(\rho)$, and $U(\mathbf{R})$) to the effective potential $U_{eff}$. Then, the configurational pressure approximately determined by the average system virial per the system volume, $p^{conf} \cong \frac{\langle W \rangle}{V} = -\frac{1}{3V}\langle \mathbf{R} \cdot \nabla U_{eff} \rangle$, and expressed as $p^{conf} = \frac{RT\rho}{M} \varphi\left(\frac{(\rho/\rho_0)^{m_{IPL}/3}}{kT}\right)$ to derive Eq. (5) in the case of the power law density scaling, can be generalized as follows

$$p^{conf} = \frac{RT\rho}{M} \varphi\left(\frac{h(\rho)}{kT}\right) \tag{8}$$

By analogy with the derivation of Eq. (5),[26,38,39,42] we perform the first order Taylor series expansion of the function $\varphi(x)$ with $x = h(\rho)/(kT)$ in Eq. (8) about $x_0 = h(\rho_0)/(kT)$, i.e., about $\rho = \rho_0$, which yields $p^{conf} = \frac{RT\rho}{M}[\varphi(x_0) + \varphi'(x_0)(x - x_0)]$. Introducing the reference configurational pressure $p_0^{conf} = \frac{RT\rho}{M}\varphi(x_0)$ and assuming the only temperature dependent parameter $B(T) = \frac{R\rho}{kM}\varphi'(x_0)$, we arrive at the following EOS

$$p^{conf} = p_0^{conf} + B(T)(h(\rho) - h(\rho_0)) \tag{9}$$

The physical meaning of the temperature dependent parameter $B(T)$ can be found by using the generalized density scaling exponent given by Eq. (2). In the isothermal conditions, density is an only pressure function $\rho = \rho(p)$. Then, we can transform Eq. (2), exploiting the configurational isothermal bulk modulus, $\frac{d\ln h(\rho)}{d\ln \rho} = \frac{\partial \ln h(\rho)}{\partial p^{conf}}\frac{\partial p^{conf}}{\partial \ln \rho}\bigg|_T = \frac{\partial \ln h(\rho)}{\partial p^{conf}}\bigg|_T B_T^{conf}$, to the following differential equation

$$\frac{\partial \ln h(\rho)}{\partial p^{conf}} = \frac{\gamma(\rho)}{B_T^{conf}(p^{conf})} \tag{10}$$

A general solution of Eq. (10) can be expressed as follows $h(\rho) = A \exp\left(\int \frac{\gamma(\rho)}{B_T^{conf}(p^{conf})} dp^{conf}\right)$. Assuming the initial condition, $\rho(T, p_0^{conf}) = \rho_0$, we find the integration constant, $A = h(\rho_0)\exp\left(-\int \frac{\gamma(\rho)}{B_T^{conf}(p^{conf})} dp^{conf}\right)\bigg|_{\rho=\rho_0}$, and the related particular solution of Eq. (10),

$$h(\rho) = h(\rho_0) \exp\left(-\int \frac{\gamma(\rho)}{B_T^{conf}(p^{conf})} dp^{conf}\right)\bigg|_{\rho=\rho_0} \exp\left(\int \frac{\gamma(\rho)}{B_T^{conf}(p^{conf})} dp^{conf}\right),$$ which can be approximated

by the following density scaling function

$$h(\rho) = h(\rho_0)\left[1 + \frac{\gamma(\rho_0)}{B_T^{conf}(p_0^{conf})}(p^{conf} - p_0^{conf})\right] \qquad (11)$$

using the first order Taylor series expansion about $p^{conf} = p_0^{conf}$,

$$\exp\left(\int \frac{\gamma(\rho)}{B_T^{conf}(p^{conf})} dp^{conf}\right) = \exp\left(\int \frac{\gamma(\rho)}{B_T^{conf}(p^{conf})} dp^{conf}\right)\bigg|_{p=p_0} \left\{1 + \frac{\gamma(\rho_0)}{B_T^{conf}(p_0^{conf})}(p^{conf} - p_0^{conf})\right\}.$$ Since

Eq. (9) implies the density scaling function, $h(\rho) = h(\rho_0) + B^{-1}(T)(p^{conf} - p_0^{conf})$ we confirm by

comparison with Eq. (11) that the parameter $B(T)$ can depend only on temperature

$$B(T) = \frac{B_T^{conf}(p_0^{conf})}{\gamma(\rho_0) h(\rho_0)} \qquad (12)$$

This is because $B_T^{conf}(p_0^{conf})$, $h(\rho_0)$, $\gamma(\rho_0)$ are determined at a constant reference pressure $p_0$. The relation given by Eq. (12) applied to Eq. (9) results in the EOS that possesses all parameters with the well-defined physical meaning,

$$p^{conf} = p_0^{conf} + \frac{B_T^{conf}(p_0^{conf})}{\gamma(\rho_0)}\left(\frac{h(\rho)}{h(\rho_0)} - 1\right) \qquad (13)$$

It should be noted that Eqs. (10) and (11) are also valid after replacing its configurational quantities with their nonconfigurational counterparts. In this way, we can formulate another EOS, which is morphologically very similar to Eq. (13),

$$p = p_0 + \frac{B_T(p_0)}{\gamma(\rho_0)}\left(\frac{h(\rho)}{h(\rho_0)} - 1\right) \qquad (14)$$

If Eqs. (13) and (14) are considered in the case of the power law density scaling, i.e., if $h(\rho) = \rho^{\gamma_{EOS}}$, then the scaling exponent is a material constant independent of thermodynamic conditions and $\gamma(\rho) = \gamma_{EOS}$ based on Eq. (2). Consequently, in this case, Eqs. (13) and (14) can be reduced to Eqs. (5) and (6), respectively. We have reported[39,40] that the important prediction made by Eqs. (5) and (6) is the linear scaling of PVT data, $(\rho/\rho_0)^{\gamma_{EOS}}$ vs. $(p-p_0)/B_T(p_0)$ or $(p^{conf}-p_0^{conf})/B_T^{conf}(p_0^{conf})$ with the slope equal to $\gamma_{EOS}$. One can note that Eqs. (13) and (14) also lead to some kind of the linear scaling of PVT data that generalizes $(\rho/\rho_0)^{\gamma_{EOS}}$ to $h(\rho)/h(\rho_0)$ and $\gamma_{EOS}$ to $\gamma(\rho_0)$, but only if $\rho_0 = const$ is chosen at each considered temperature at a reference pressure $p_0$. It means that the slope of the linear dependences $h(\rho)/h(\rho_0)$ vs. $(p-p_0)/B_T(p_0)$ or $(p^{conf}-p_0^{conf})/B_T^{conf}(p_0^{conf})$ possesses a particular value, which depends on the choice of the reference state and does not always constitute any representative value of the scaling exponent γ if $\gamma(\rho)$ changes with density.

In addition, it is interesting to determine the pressure dependences of the configurational isothermal bulk modulus $B_T^{conf}(p)$ and the isothermal bulk modulus $B_T(p)$ that follow from Eqs. (13) and (14) to compare them with the mentioned linear pressure functions $B_T^{conf}(p)$ and $B_T(p)$ implied by Eqs. (5) and (6), respectively. Differentiating the equations of state Eqs. (13) and (14) with respect to $\ln \rho$ and exploiting Eq. (2), we find the following relations

$$B_T^{conf}(p) = B_T^{conf}(p_0^{conf}) \frac{\gamma(\rho)h(\rho)}{\gamma(\rho_0)h(\rho_0)} \tag{15}$$

$$B_T(p) = B_T(p_0) \frac{\gamma(\rho)h(\rho)}{\gamma(\rho_0)h(\rho_0)} \tag{16}$$

which are reasonable results because the values of the fitting parameters involved in the functions $h(\rho)$ and $\gamma(\rho)$ can be slightly different in Eqs. (13) and (14). If $h(\rho) = \rho^{\gamma_{EOS}}$, then $\gamma(\rho) = \gamma(\rho_0) = \gamma_{EOS}$ and $h(\rho)/h(\rho_0) = (\rho/\rho_0)^{\gamma_{EOS}}$. If we replace the right side of the latter equation with its counterparts found from Eqs. (5) and (6), respectively, we indeed arrive at the mentioned linear pressure dependences $B_T^{conf}(p)$ and $B_T(p)$. However, in general, Eqs. (15) and (16) can result in nonlinear pressure functions in isothermal conditions. Exploiting Eq. (11) and its counterpart for nonconfigurational quantities, we find that the pressure dependences of $B_T^{conf}$ and $B_T$ can deviate from the linear character in isothermal conditions if the scaling exponent γ varies with density,

$$B_T^{conf}(p) = \frac{\gamma(\rho)}{\gamma(\rho_0)}\left[B_T^{conf}(p_0^{conf}) + \gamma(\rho_0)(p^{conf} - p_0^{conf})\right] \tag{17}$$

$$B_T(p) = \frac{\gamma(\rho)}{\gamma(\rho_0)}\left[B_T(p_0) + \gamma(\rho_0)(p - p_0)\right] \tag{18}$$

**III. TEST OF THE EOS BY USING A PROTOTYPICAL MODEL OF SUPERCOOLED LIQUID**

In Section II, we have derived two morphologically similar equations of state, which are based on the assumption that the scaling exponent can be a function $\gamma(\rho)$, dependent only on density and related by Eq. (2) and the theory of isomorphs to a density scaling function $h(\rho)$ not limited to a power density function. Based on the theory of isomorphs, one can also show[31,34] that the only density dependent function $h$ in Eq. (1) is given by the polynomial $h(\rho) = \sum_j c_j \rho^{m_j/3}$ in the case of a *strongly correlating system*, the molecular dynamics of which is described by the multiple IPL potential defined by a sum of the IPL terms $\sum_j u_j r^{-m_j}$. Therefore, Eqs. (13) and (14) with the density scaling function, $h(\rho) = \sum_j c_j \rho^{m_j/3}$, can be applied to describe PVT data of systems characterized by such an IPL potential. For instance, the following density scaling function

$$h(\rho) = c\rho^4 + (1-c)\rho^2 \tag{19}$$

has been argued[31,34] for the KABLJ model based on the LJ potential with the exponents of the repulsive and attractive terms equal to 12 and 6, respectively. Then, Eq. (2) implies the related density function for the scaling exponent

$$\gamma(\rho) = \frac{4c\rho^4 + 2(1-c)\rho^2}{h(\rho)} \tag{20}$$

Thus, Eqs. (13) and (14) with the functions $h(\rho)$ and $\gamma(\rho)$ given respectively by Eqs. (19) and (20) should be valid in the case of the KABLJ model.

It is worth noting that a known function $h(\rho)$ with values of its parameters found from fitting PVT data to the EOS given by Eqs. (13) or (14) is expected to enable to scale the structural relaxation times according to Eq. (1). The suggested property is of great importance to the proper linkage between dynamics and thermodynamics near the glass transition, which is still under investigation. Therefore, besides the standard test for a good approximation of the PVT data by Eqs. (13) and (14), we also verify whether these EOS meet this important criterion or not.

To perform the test we exploit our MD simulation data collected from the equilibrium simulation of 1000 particles in the KABLJ model in the NVT ensemble. Some of the used isotherms of structural relaxation times and PVT data (at $T$=0.5, 0.75, 1.0, 1.5, 2.0, 2.5, 3.0 in the LJ units) have been earlier reported.[26] However, the related particle number density range, $1.2 \leq \rho \leq 1.6$ in LJ units, is then too narrow to be representative for the general density scaling regime not limited to the power law density scaling. As already mentioned in Introduction, the power law density scaling of structural relaxation times $\tau$ in the KABLJ model can be achieved[30,31] when the particle number density ranges from 1.2 to 1.6 if the reduced units suggested by Dyre's group are employed (e.g., $\tilde{\tau} = \tau \rho^{1/3} T^{1/2}$, for $\tau$, $\rho$, $T$ in the usual LJ units) to cause the NVT molecular dynamics to be isomorph invariant in the sense postulated by the theory of isomorphs.[29] Nevertheless, the power law density scaling of structural relaxation times is impossible in the considerably wider range of particle number densities, i.e., $1.2 \leq \rho \leq 2.0$, even if the structural relaxation times are expressed in the reduced units.[31] Therefore, we have performed additional MD simulations in the KABLJ model at temperatures $T$=2.0, 3.0, 3.5, 4.0, 4.5, 5.0 using the RUMD package[44] to supplement the earlier collected simulation data with the particle number density range, $1.6 < \rho \leq 2.0$. Other details of the simulations are the same as those described in Ref. 26 for the narrower particle number density range, $1.2 \leq \rho \leq 1.6$.

The structural relaxation times have been determined in the usual manner[30] from incoherent intermediate self-scattering functions[27] ($\tau = t$ if $F_S(\mathbf{q},t) = e^{-1}$) at the wave vector $\mathbf{q}$ of the first peak of the AA structure factor particles at each simulation state ($T,\rho$) separately, where A denotes the particle specie that constitutes 80% of the particle content of the binary mixture. As can be seen in Fig. 1, there are only small differences in $\tau$ in the LJ units (Fig. 1(a)) and $\tilde{\tau}$ in the reduced units (Fig. 1(b)) suggested by the theory of isomorphs. However, we perform the further analysis for the KABLJ model in the reduced units to meet the requirements of the theory of isomorphs. In this context, it should be noted that the quotient character of Eqs. (13) and (14) allows us to apply these EOS to describe PVT data in the usual LJ units.

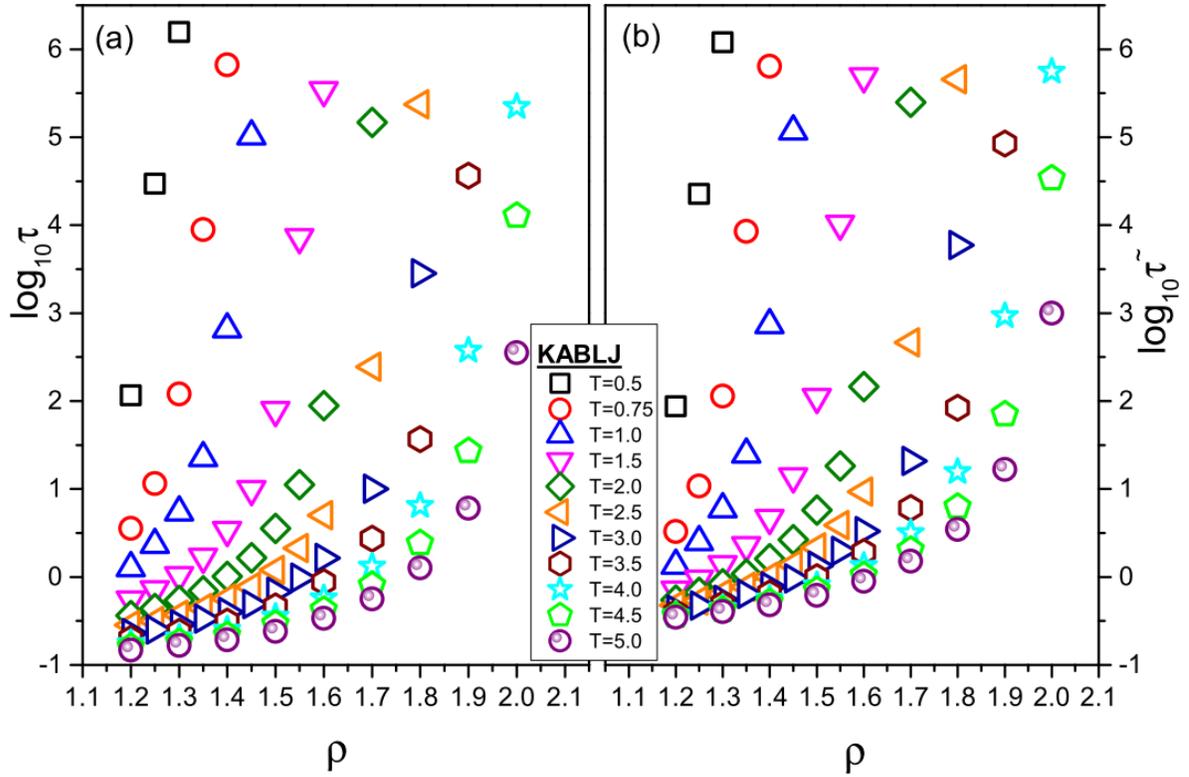

FIG. 1. Plot of isothermal structural relaxation times in the KABLJ model at different temperatures versus the particle number density: (a) $\tau$ in the usual LJ umits and (b) $\tilde{\tau}$ in the reduced units, $\tilde{\tau} = \tau\rho^{1/3}T^{1/2}$, where the particle number density ρ and temperature T are in the usual LJ units.

It is worth noting that the density dependent functions $h(\rho)$ and $\gamma(\rho)$ given by Eqs. (19) and (20) have only one shared parameter $c$, which means that the number of fitting parameters in the generalized density scaling isothermal EOS given by Eqs. (13) and (14) does not increase in comparison with their power law density scaling counterparts represented by the isothermal versions of Eqs. (5) and (6) that are characterized by the power density function $h(\rho) = \rho^{\gamma_{EOS}}$ and the constant scaling exponent $\gamma(\rho) = \gamma_{EOS}$. Using the simulation data in the KABLJ model in the particle number density range, $1.2 \leq \rho \leq 1.6$ in the usual LJ units, we have previously shown[26] that the scaling exponent γ that enables the power law density scaling of structural relaxation times of the KABLJ liquid is approximately the same as those found from fitting PVT data of the KABLJ model to Eqs. (5) or (6). Now, we check whether it is possible to find the value of the parameter $c$ from fitting PVT data collected for MD simulations in the KABLJ model in the considerably larger particle number density range, $1.2 \leq \rho \leq 2.0$ in the usual LJ units, which also enables the function $h(\rho)$ given by Eq. (19) to scale the structural relaxation times according to Eq. (1).

The polynomial character of the functions $h(\rho)$ given by Eq. (19) that also affects the form of the function $\gamma(\rho)$ given by Eq. (20) causes that it is convenient to use reduced densities $\rho^{reduc} = \rho/\rho_{ref}$ to ensure that the values of the parameters $c$ and $(1-c)$ have consistent units. This fact and an intrinsic feature of the theory of isomorphs, which predicts (see Eq. (2) in Ref. 31) a micro-configuration (all particle coordinates) $(\rho/\rho_{ref})^{-1/3}\mathbf{R}$ at a state $(T,\rho)$ isomorphic with the chosen reference state $(T_{ref},\rho_{ref})$ described by the micro-configuration $\mathbf{R}$, cause that we exploit reduced densities to verify the new equations of state (Eqs. (13) and (14)) with the density functions Eqs. (19) and (20) as well as to scale structural relaxation times according to Eq. (1). Nevertheless, the plots of PVT data in the KABLJ model with their fitting curves to Eqs. (13) and (14) are shown (see Fig. 2) as pressure functions of non-reduced density for convenience of reading the figures.

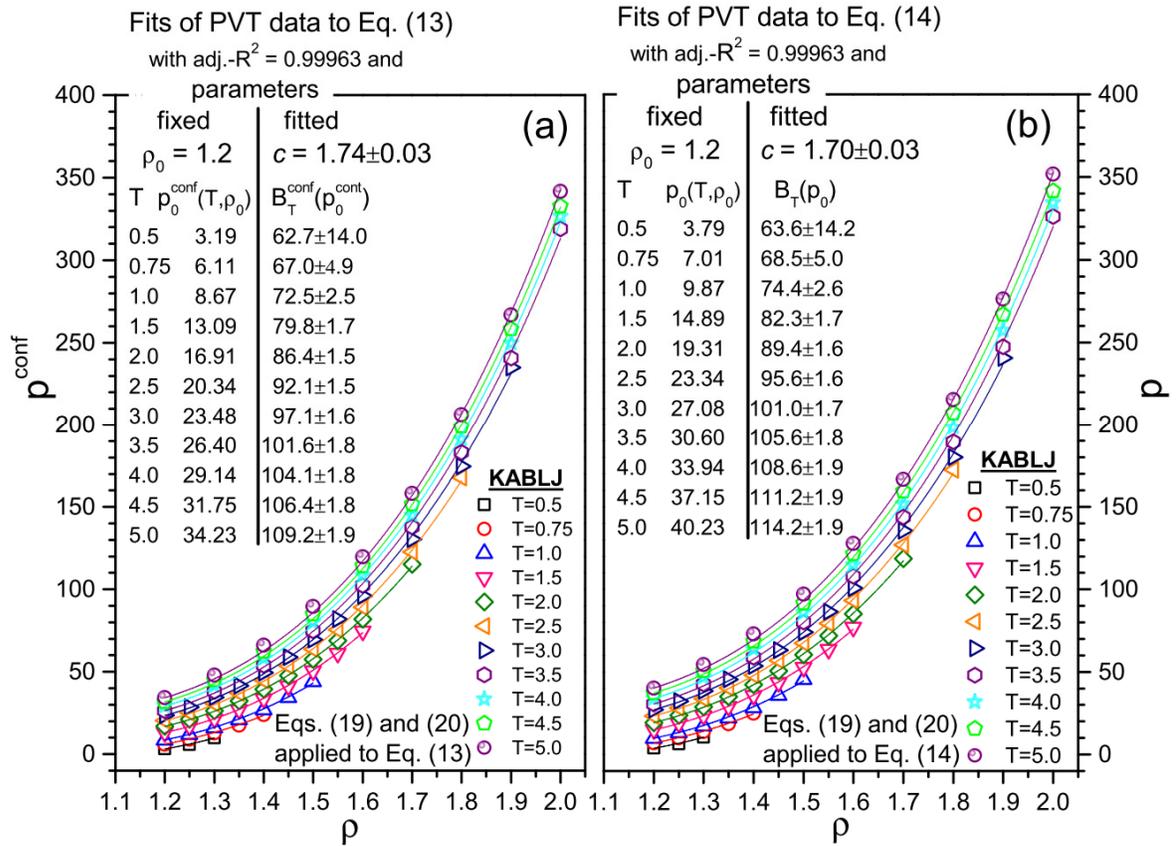

FIG. 2. Plot of isothermal volumetric data in the KABLJ model at different temperatures: (a) the configurational pressure versus the particle number density and (b) pressure versus the particle number density. Values of all quantities are in the usuall LJ units. The solid lines in panels (a) and (b) denote fitting curves of the PVT data respectively to Eq. (13) and Eq. (14) with the density dependent functions $h(\rho)$ and $\gamma(\rho)$ given by Eqs. (19) and (20), where the normalized densities $\rho^{norm} = \rho/\rho_{high}$ with $\rho_{high} = 2.0$ in the usual LJ unit are assumed instead of $\rho$.

It is reasonable to assume $\rho_0 = 1.2$ to fit the PVT data to Eqs. (13) and (14), that is the lowest density at which the values of the reference pressure are known for each isotherm of tested PVT simulation data, and consequently $\rho_{ref} = \rho_0$ to calculate the reduced densities $\rho^{reduc} = \rho/\rho_{ref}$ as $\rho^{reduc} = \rho/\rho_0$. Then, based on Eqs. (19) and (20), $h(\rho_0^{reduc}) = h(1) = 1$ and $\gamma(\rho_0^{reduc}) = \gamma(1) = 2c + 2$. The latter linear equation is a special case of Eq. (20), which in general results in a rational nonlinear function of the parameter $c$ at a given density. To make the fitting procedure more reliable, it is better to avoid this non-representative linear case, especially that the fitting value of the parameter $c$ tends then to infinity for numerical reasons. In order to do that we normalize the explored density domain $1.2 \leq \rho \leq 2.0$ in the usual LJ units to the range, $0.6 \leq \rho^{norm} \leq 1.0$, where $\rho^{norm} = \rho/\rho_{high}$ with the highest considered density $\rho_{high} = 2.0$ in the usual LJ units. Using the normalized density domain, we are able to find reliable values of the fitting parameter $c$ as well as $B_T^{conf}(p_0^{conf})$ and $B_T(p_0)$ respectively for Eqs. (13) and (14) with the functions $h(\rho)$ and $\gamma(\rho)$ given by Eqs. (19) and (20). It should be stressed that the values of configurational reference pressures $p_0^{conf}$ and reference pressures $p_0$ have not been fitted, because they have been fixed in the case of each isotherm based on the simulation PVT data at $\rho_0 = 1.2$ in the usual LJ units. The fitting curves of the PVT simulation data to Eqs. (13) and (14) with their fitting values of the parameters are presented in Figs. 2(a) and 2(b), respectively. As a result, we obtain a good quality of the fits and very close values of the parameter $c = 1.74 \pm 0.03$ and $c = 1.70 \pm 0.03$ determined from Eqs. (13) and (14), respectively, using the functions $h(\rho)$ and $\gamma(\rho)$ given by Eqs. (19) and (20). Then, we apply the function $h(\rho)$ defined by Eq. (19) with both the established values of the parameter $c$ to scale the structural relaxation times of the KABLJ model in the reduced units (i.e., $\tilde{\tau} = \tau \rho^{1/3} T^{1/2}$) according to Eq. (1). To do that we exploit the assumed reduced density, $\rho^{reduc} = \rho/\rho_0$, that implies the scaling function $h(\rho/\rho_0) = c(\rho/\rho_0)^4 + (1-c)(\rho/\rho_0)^2$, which successfully leads (see Figs. 3(a) and 3(b)) to the generalized density scaling of the structural relaxation times $\tilde{\tau}$ in terms of Eq. (1) for both the values of the parameter $c$. It means that the generalized density scaling should be also possible in the normalized density range, because the function $h(\rho/\rho_0)$ results in $h(\rho^{norm}/\rho_0^{norm}) = c(\rho_{high}/\rho_0)^4(\rho/\rho_{high})^4 + (1-c)(\rho_{high}/\rho_0)^2(\rho/\rho_{high})^2$, but to scale the structural relaxation times by means of the function, $h(\rho^{norm}/\rho_0^{norm}) = h(\rho/\rho_{high})$, we need to rescale the polynomial coefficients $c$ and $(1-c)$ to $c(\rho_{high}/\rho_0)^4$ and $(1-c)(\rho_{high}/\rho_0)^2$, respectively.

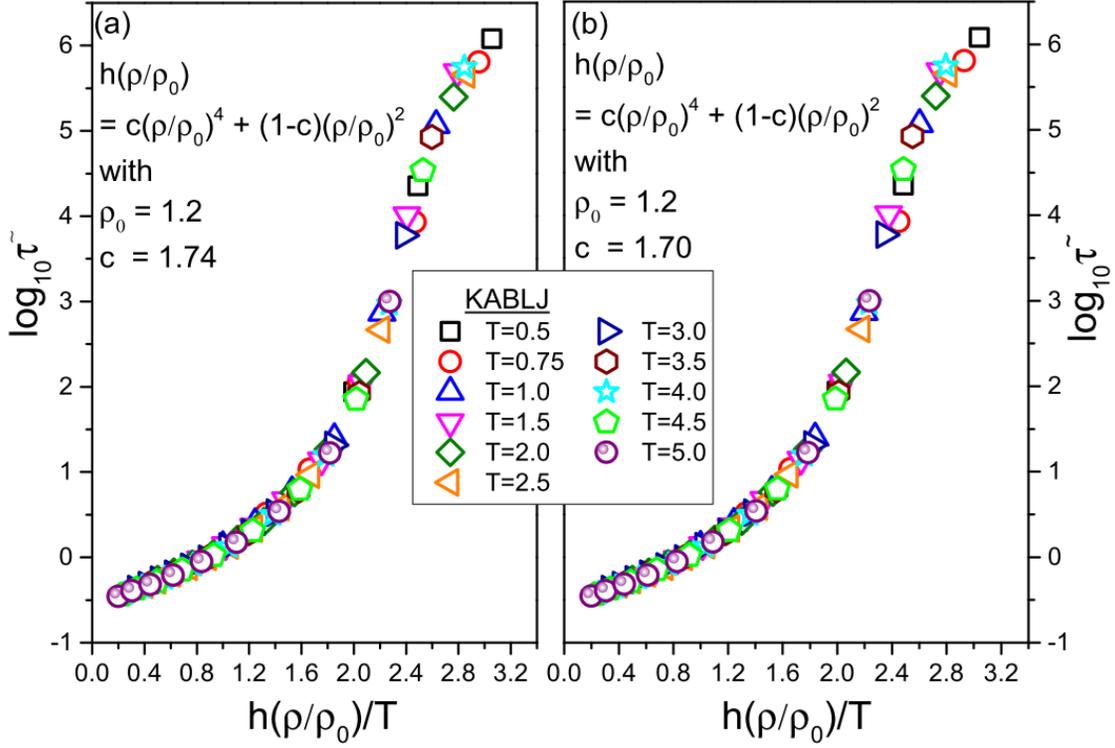

FIG. 3. The density scaling of all the isotherms of the structural relaxation times in the reduced units in the KABLJ model from Fig. 1(b) according to Eq. (1) with the density scaling functions given by Eq. (19) with the value of the parameter $c$ found from fitting PVT data to the equation of state given respectively by (a) Eq. (13) and (b) Eq. (14).

As already mentioned and discussed in the context of Eq. (2), the scaling exponent $\gamma$ in the case of the generalized density scaling depends on density. It is interesting to see how varies the density dependent exponent $\gamma(\rho)$ given by Eq. (20) in the KABLJ model examined in the density range $1.2 \leq \rho \leq 2.0$ in the usual LJ units and how the changes in $\gamma(\rho)$ affect the pressure dependences $B_T^{conf}(p)$ and $B_T(p)$, which are predicted respectively by Eqs. (17) and (18) to deviate from the linear character valid at low densities or pressures. It is worth noting that the PVT data are considered herein in two times larger density range ($1.2 \leq \rho \leq 2.0$ in the usual LJ units) that corresponds to three and a half times larger pressure range ($0 < p \leq 350$ in the usual LJ units) than those ($1.2 \leq \rho \leq 1.6$, $0 < p \leq 100$ in the usual LJ units) explored by us previously[26] to test Eqs. (5) and (6) by means of the KABLJ model. We found that the increase in density results in a decrease in the value of the scaling exponent from $\gamma = 5.48$ at $\rho = 1.2$ in the usual LJ units to $\gamma = 4.44$ at $\rho = 2.0$ in the usual LJ units if we calculate the values $\gamma(\rho/\rho_0)$ from Eq. (20) with $c = 1.70$ obtained from fitting the PVT data to Eq. (13), and in a very slightly different decrease in these value from $\gamma = 5.40$ at $\rho = 1.2$ in the usual LJ units to $\gamma = 4.35$ at $\rho = 2.0$ in the usual LJ units if we calculate the values $\gamma(\rho/\rho_0)$ from Eq. (20) with $c = 1.70$ obtained from fitting the PVT data to Eq. (14). These changes in γ with varying density should influence the pressure dependences of the configurational isothermal bulk modulus and the isothermal bulk modulus according to Eqs. (17) and (18), respectively. To examine the high pressure behavior of $B_T^{conf}$ and $B_T$, we choose the PVT isotherms at high temperatures, because the PVT data simulated for these isotherms cover the entire considered density range $1.2 \leq \rho \leq 2.0$ in the usual LJ units. As an example, we show the dependences $B_T^{conf}(p)$ and $B_T(p)$ at T=3.5 in Fig. 4.

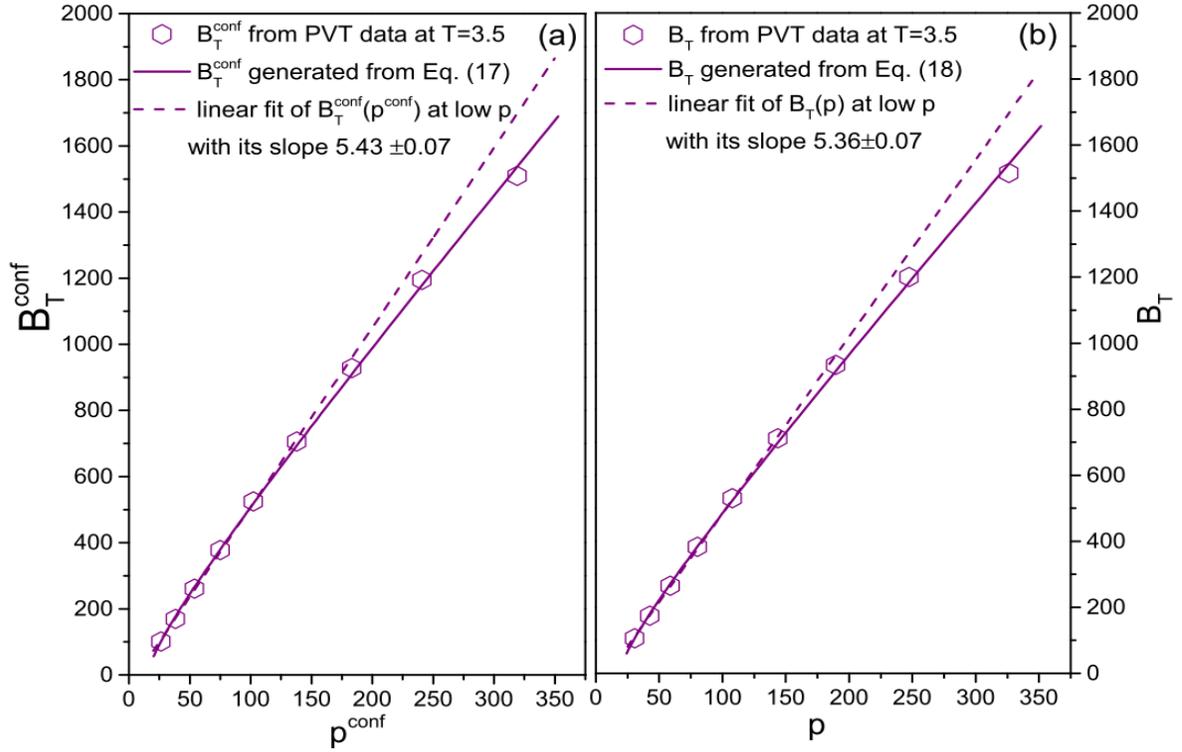

FIG. 4. Plots of the configurational isothermal bulk modulus versus the configurational pressure (a) and (b) the isothermal bulk modulus versus pressure calculated respectively from the definitions $B_T^{conf} = (\partial p^{conf}/\partial \ln \rho)_T$ and $B_T = (\partial p/\partial \ln \rho)_T$ by using the PVT simulation data in the KABLJ model at T=3.5 in the usual LJ units. The dashed lines denote the low pressure linear fits of these dependences and their high pressure extrapolations. The solid lines are generated in panels (a) and (b) respectively from Eq. (17) and Eq. (18) with the values of their paramaters found by fitting the PVT data (see Fig. 2) respectively to Eqs. (13) and (14) with the density dependent functions $h(\rho)$ and $\gamma(\rho)$ given by Eqs. (19) and (20), exploiting the normalized densities $\rho^{norm} = \rho/\rho_{high}$ with $\rho_{high} = 2.0$ in the usual LJ unit instead of ρ as it was assumed to fit the PVT data to these equations of state.

The expected deviation of these dependences from the linear character is indeed observed at pressures higher than 100 in the usual LJ units. For instance, the value of $B_T$ at the highest simulated pressure along the isotherm $T=3.5$ is 12% smaller than its counterpart calculated from the high pressure extrapolation of the linear pressure dependence of $B_T$ determined at low pressures with its slope 5.36, which is very close to the mentioned low density limit ($\gamma$=5.40 at $\rho=1.2$) of the density function of the scaling exponent $\gamma$.

## IV. THE SUGGESTED FORM OF THE EOS FOR PVT DATA OF REAL GLASS FORMERS

Very recently, another density scaling function $h(\rho)$ established phenomenologically

$$h(\rho) = \exp\left(C_1 \ln \rho + C_2 \ln^2 \rho\right) \qquad (21)$$

has been suggested[31] to scale structural relaxation times of real glass formers in terms of Eq. (1). For the convenience of calculations, the polynomial function of $\ln \rho$ is used here in Eq. (21) instead of the polynomial function of $\log_{10} \rho$, which means that the fitting parameters $C_1 = A_1$ and $C_2 = A_2/\ln 10$ if $A_1$ and $A_2$ are the fitting parameters of the function $h(\rho)$ reported for real glass formers in Ref. 31. It should be noted that Eq. (21) via Eq. (2) implies the following simple density dependent function for the scaling exponent

$$\gamma(\rho) = C_1 + 2C_2 \ln \rho \qquad (22)$$

which can be reduced to the constant scaling exponent if the parameter $C_2 = 0$. However, the parameter $C_1$ indicates in general the value of the scaling exponent $\gamma(\rho)$ if density tends to unity. Taking into consideration the successful description of the deviation from the power law density scaling in the case of supercooled van der Waals liquids measured in a wide pressure range (i.e., in a wide density range), which is possible by using Eq. (21),[31] we postulate that Eqs. (21) and (22) cause the equations of state (Eqs. (13) and (14)) to properly describe PVT data of real glass formers, the molecular dynamics of which obeys the generalized density scaling law (Eq. (1)), because these EOS are derived herein also in the density scaling regime not limited to the power law density scaling.

Unfortunately, in contrast to the KABLJ model successfully used to verify the derived EOS with the functions $h(\rho)$ and $\gamma(\rho)$ given by Eqs. (19) and (20), a reliable test of Eqs. (13) and (14) with the density functions $h(\rho)$ and $\gamma(\rho)$ given by Eqs. (21) and (22) can be only done indirectly due to the high pressure limit of PVT measurements, which usually do not exceed 200MPa. Although an exception is made by the PVT experimental data of a few glass formers measured up to 700MPa,[8] the very recent ultra-high pressure dielectric and ultrasonic measurements of propylene carbonate (PC) up to 4.2GPa and 1.7GPa, respectively, show[45] that the pressure dependence of the isothermal bulk modulus determined from the ultrasonic measurements within the pressure range from 0.1MPa to about 1GPa at T=295K, which is above the glass transition temperature of PC, cannot be satisfactorily described in the low pressure limit by using the equation of state based on the Tait equation, which has been parameterized for PC in Ref. 8. On the other hand, the Tait equation earlier parameterized by using PVT data of PC[6] measured in the typical pressure range of PVT experiments, i.e., up to 200MPa, properly predicts the low pressure limit of the dependence $B_T(p)$ found from the ultrasonic data, but these dependences begin to diverge with increasing pressure. In this situation, Eqs. (13) and (14) with the functions $h(\rho)$ and $\gamma(\rho)$ given by Eqs. (21) and (22) seem to be a good alternative to properly describe PVT data in the ultra-high pressure limit. A preliminary test for the validity of these EOS can be based on the description of the pressure dependence of the isothermal bulk modulus of PC at T=295K, which has been reported in Fig. 8(a) in Ref. 45 for different ultrasonic experimental methods. We consider herein (Fig. 5) only the data from autocorrelation measurement assessed by the authors of Ref. 45 as the most accurate one.

It should be noted that the simple density scaling function given by Eq. (21) enables us to find pressure functions of the configurational isothermal bulk modulus and the isothermal bulk modulus based on the equations of state Eqs. (13) and (14), respectively. Since we test only the pressure dependence of the isothermal bulk modulus based on the data taken from Ref. 45, we discuss in detail only the derivation of the function $B_T(p)$. From Eq. (14), one can easily find, $h(\rho) = h(\rho_0)[1 + \gamma(\rho_0)(p - p_0)/B_T(p_0)]$. If we substitute the function $h(\rho)$ for Eq. (21) in the latter equation, we can formulate a quadratic equation for $\ln \rho$ with the following coefficients $C_2$, $C_1$, and $-(C_2 \ln^2 \rho_0 + C_1 \ln \rho_0 + \ln[1 + \gamma(\rho_0)(p - p_0)/B_T(p_0)])$. Then, we can find density as a function of pressure at a given temperature

$$\ln \rho = \frac{-C_1 \pm (C_1^2 + 4C_2^2 \ln^2 \rho_0 + 4C_2 C_1 \ln \rho_0 + 4C_2 \ln[1 + \gamma(\rho_0)(p - p_0)/B_T(p_0)])^{1/2}}{2C_2} \quad (23)$$

which can be simplified by using the reduced density $\rho^{reduc} = \rho/\rho_0$ and assuming $\rho^{reduc} \geq 1$ in the following way $\ln \rho^{reduc} = \frac{-C_1 + (C_1^2 + 4C_2 \ln[1 + \gamma(1)(p - p_0)/B_T(p_0)])^{1/2}}{2C_2}$ , where $\gamma(1) = C_1$. It is worth noting that Eq. (23) applied to Eq. (22) also enables us to establish the scaling exponent γ as a function of pressure, $\gamma(p) = \left(C_1^2 + 4C_2 \ln[1 + C_1(p - p_0)/B_T(p_0)]\right)^{1/2}$, where the reduced density is employed. Finally, Eq. (18) with the scaling exponent γ given by Eq. (22), which is expressed by the pressure function $\gamma(p)$ that assumes $\rho^{reduc} = \rho/\rho_0$, yields the following pressure function for the isothermal bulk modulus

$$B_T(p) = \frac{\left(C_1^2 + 4C_2 \ln[1 + C_1(p - p_0)/B_T(p_0)]\right)^{1/2} [B_T(p_0) + C_1(p - p_0)]}{C_1} \quad (24)$$

The authors of Ref. 45 noted that the dependence $B_T(p)$ obtained from the autocorrelation ultrasonic experiment at $T$=295K is almost linear in the considered pressure range up to 1GPa, although they observed that the isothermal compression of PC cause the dependence $B_T(p)$ to slightly deviate from the linear behavior. An interesting question arises as to how strong is the effect of pressure on the isothermal compressibility of the material squeezed above 1GPa. Assuming a gradual pressure influence on volumetric properties of supercooled liquids, one can expect that the dependence $B_T(p)$ at $p$ > 1GPa also gradually deviates from the high pressure extrapolation of the linear dependence $B_T(p)$ determined at low pressures to pressures larger than 1GPa. Taking into account this supposition, we can discriminate between two predictions about the isothermal dependence $B_T(p)$ at pressures above 1GPa, which are based on the typical quadratic pressure parametrization $B_T(p) = B_T(p_0) + (\partial B_T/\partial p)_{p=p_0}(p - p_0) + (1/2)(\partial^2 B_T/\partial p^2)_{p=p_0}(p - p_0)^2$ as well as on Eq. (24) that follows from Eq. (14) with the functions $h(\rho)$ and $\gamma(\rho)$ given by Eqs. (21) and (22). Since $B_T(p_0) = 2716.7 \text{MPa}$ can be taken from the experimental autocorrelation ultrasonic data at ambient pressure $p_0$, both the quadratic parameterization of $B_T(p)$ and Eq. (24) have two fitting parameters $D_1 = (\partial B_T/\partial p)_{p=p_0}$ and $D_2 = (1/2)(\partial^2 B_T/\partial p^2)_{p=p_0}$ as well as $C_1$ and $C_2$, respectively. We determine the values of the parameters ($D_1 = 8.8 \pm 0.1$ and $D_2 = -0.83 \pm 0.10 \text{GPa}^{-1}$ as well as $C_1 = 1.65 \pm 0.02$ and $C_2 = 7.15 \pm 0.03$) by fitting the experimental dependence $B_T(p)$ for PC at $T$=295 (see Fig. 5). Then, we extrapolate the quadratic parameterization of $B_T(p)$ and Eq. (24) with the values of their parameters fitted within the pressure range below 1GPa up to $p$=4GPa. In this way, we investigate the extremely wide pressure range that becomes achievable experimentally by means of modern high pressure techniques, e.g. exploiting diamond anvil pressure cells.

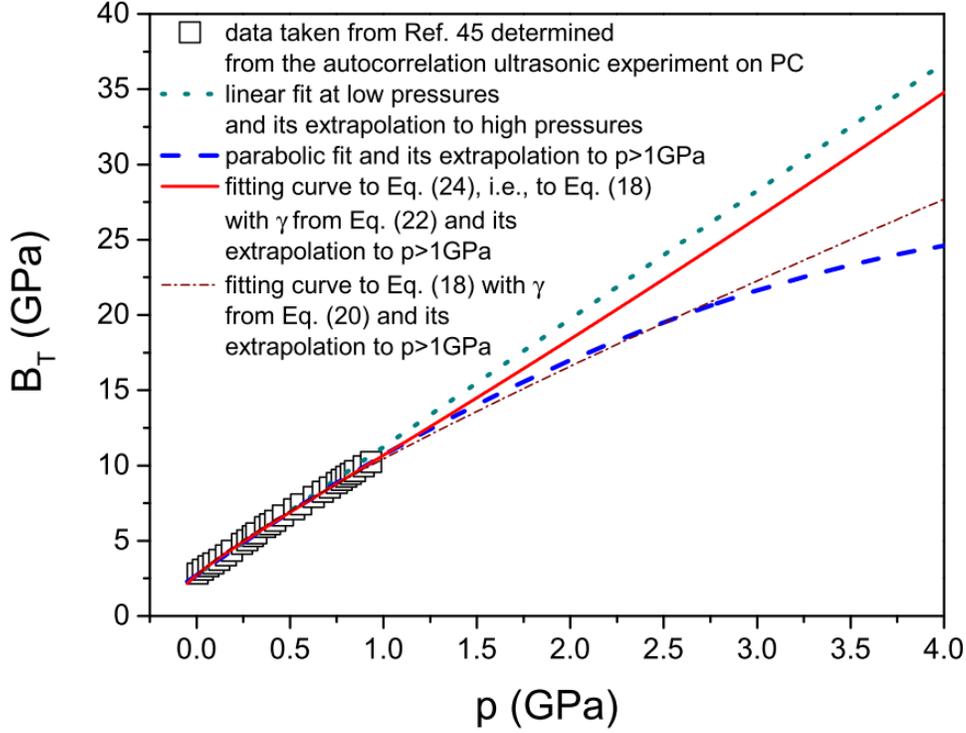

FIG. 5. A comparison of three ultra-high pressure predictions of the pressure dependence of the isothermal bulk modulus (open squares) established[45] from the ultrasonic autocorrelation measurements of PC at $T$=295K (above the glass transition temperature of PC) and pressures lower than 1GPa. The dotted line denotes the linear pressure function fitted by using the experimental data $B_T(p)$. The solid line is the fitting curve of the experimental data $B_T(p)$ to Eq. (24), i.e, to Eq. (18) with the function γ from Eq. (22), which is based on the density scaling function $h$ from Eq. (21). The dot dashed line is the fitting curve of the experimental data $B_T(p)$ to Eq. (18) with the function γ from Eq. (20), which is based on the density scaling function $h$ from Eq. (19). The dotted and dashed lines result respectively from the linear and quadratic pressure parameterizations of the experimental dependence $B_T(p)$. Each fitting curve is extrapolated up to $p$=4GPa.

As a result, we find that the high pressure extrapolation based on the quadratic parameterization of $B_T(p)$, denoted by the dashed line in Fig. 5, diverges from the linear behavior (the dotted line in Fig. 5) in a significantly larger degree than that established for the high pressure extrapolation made by means of Eq. (24), depicted by the solid line in Fig. 5. For instance, the values of $B_T(p)$ extrapolated at $p$=4GPa by using the quadratic pressure parameterization and Eq. (24) are respectively 45% and 6% less than that predicted by the linear pressure dependence $B_T(p)$ found at low pressures. This result indirectly gives evidence of the good ability of Eq. (14) with the functions $h(\rho)$ and $\gamma(\rho)$ assumed by Eqs. (21) and (22) to describe the ultra-high pressure PVT data. Nevertheless, this preliminary test requires verifying in the future by using the ultra-high experimental PVT data or at least the pressure dependences of the isothermal bulk modulus measured above 1GPa at different temperatures. A similar stipulation can be made in the case of Eq. (23). As already mentioned, Eq. (23) provides a pressure function for density, which could be used to evaluate density in the ultra-high pressure range if the values of the parameters $C_1$ and $C_2$ are known. To do that one could employ the temperature parameterizations of the density $\rho_0 = \rho(T, p_0)$ and the isothermal bulk modulus $B_T(p_0)$ at the reference pressure $p_0$ exploited by us in Eqs. (5) and (6). Then, one could establish the temperature-density dependence of structural relaxation times and make an attempt at scaling them according to Eq. (1). However, to make such an analysis more reliable, the found values of the parameters $C_1$ and $C_2$ should be prior verified by using the ultra-high pressure experimental PVT data or at least a larger dataset of the pressure dependences of the isothermal bulk modulus measured from ambient pressure to the highest pressure as possible (preferably $p>1$GPa) at different temperatures within the explored temperature range.

Finally, it is interesting to discuss the ultra-high pressure prediction of the dependence $B_T(p)$ for PC at $T=295$, which results from Eq. (18) with the density function $\gamma(\rho)$ given by Eq. (20) that is a consequence (via Eq. (2) with Eq. (19)) of the LJ potential with the exponents 12 and 6 of the repulsive and attractive terms, respectively. In this case, using Eq. (14) with the functions $h(\rho)$ and $\gamma(\rho)$ given by Eqs. (19) and (20), we determine, $\rho^2 = \dfrac{-(1-c) \pm \left((1-c)^2 + 4ch(\rho_0)[1+\gamma(\rho_0)(p-p_0)/B_T(p_0)]\right)^{1/2}}{2c}$, which is next appropriately employed in Eq. (18) to find the pressure function of $B_T$. Assuming the reference density $B_T(p_0) = 2716.7$MPa and $\rho_0 = 1.19$g/cm$^3$ at ambient pressure,[45] the value of only one parameter $c$ requires estimating. By fitting the ultrasonic autocorrelation data for the isothermal bulk modulus of PC at $p<1$GPa to Eq. (18) with the density function $\gamma(\rho)$ given by Eq. (20), we obtain $c = 2.23 \pm 0.01$, where the normalized density domain defined by the transformation, $\rho^{norm} = \rho/\rho_{high}$, with the arbitrarily chosen sufficiently large density, $\rho_{high} = 2.0$g/cm$^3$, has turned out to be better to use for the same reasons as those pointed out in the case of the fitting of the PVT simulation data in the KABLJ model to Eqs. (13) and (14) with the functions $h(\rho)$ and $\gamma(\rho)$ given by Eqs. (19) and (20). Then, we extrapolate this fitting curve of the dependence $B_T(p)$ to higher pressures up to $p=4$GPa. This prediction (depicted by the dot dashed line in Fig. 5) is considerably closer to that established by using the quadratic pressure parameterization of $B_T$ (the dashed line in Fig. 5) than it is relative to that determined from Eq. (24), which is denoted by the solid line in Fig. 5. It means that we obtain the significantly different predictions of the dependence $B_T(p)$ above 1GPa from Eq. (18) with the scaling exponent $\gamma(\rho)$ calculated from the density scaling functions $h(\rho)$ given by Eqs. (21) and (19), for which real glass formers have been previously found[31] to meet and not to meet the generalized density scaling criterion (Eq. (4)), respectively. Thus, it is reasonable to claim that the high pressure behavior of the isothermal bulk modulus can reflect the ability of a given function $h(\rho)$ to scale the molecular dynamics near the glass transition according to the generalized density scaling law given by Eq. (1) and *vice versa* the scaling function $h(\rho)$ can be used to predict at least approximately the pressure dependence of the isothermal bulk modulus.

**V. SUMMARY AND CONCLUSIONS**

We have generalized two equations of state (Eqs. (5) and (6)) earlier discussed[26,38,39,40] in the power law density scaling regime to describe PVT data of supercooled liquids, the molecular dynamics of which obeys the density scaling law not limited to the power law density scaling. The generalization about Eq. (5) is Eq. (13) derived by using the effective intermolecular potential (Eq. (7)) with well-separated density and configuration contributions as well as the general relation (Eq. (2)) argued by the theory of isomorphs between the density dependent scaling exponent $\gamma(\rho)$ and the density scaling function $h(\rho)$, which has been combined with the definition of the configurational isothermal bulk modulus. However, the generalization about Eq. (6) is Eq. (14) formulated from Eq. (2) combined with the definition of the isothermal bulk modulus.

We have very successfully verified both these EOS by using the PVT data obtained from MD NVT simulations in the KABLJ model in the extremely wide density range ($1.2 \leq \rho \leq 2.0$ in the usual LJ units), which corresponds to the extremely wide pressure range ($0 < p \leq 350$ in the usual LJ units). To perform this test the functions $h(\rho)$ and $\gamma(\rho)$ in Eqs. (13) and (14) have been given by Eqs. (19) and (20), which can be analytically derived[31,34] for the intermolecular potential consisted of two IPL terms. By fitting the PVT data from MD simulations in the KABLJ model to each equation of state with the functions $h(\rho)$ and $\gamma(\rho)$ with only one shared parameter, we have found the value of this parameter, and using this value in Eq. (1) with the functions $h(\rho)$ given by Eq. (19), we have scaled the structural relaxation times evaluated in the extremely wide density range ($1.2 \leq \rho \leq 2.0$ in the usual LJ units) in the KABLJ model. This result obtained for each EOS is meaningful, because it confirms the validity of the derived equations of state in the generalized density scaling regime. This scaling has also revealed a peculiar behavior of the longest structural relaxation times, which seem to increase considerably slower than the shorter relaxation times do. However, this observation requires verifying by extremely time-consuming MD simulations, the results of which would be able to exceed as much as possible the structural relaxation times currently reached.

Based on the generalized equations of state given by Eqs. (13) and (14), we have arrived at the pressure dependences of the configurational isothermal bulk modulus (Eq. (17)) and the isothermal bulk modulus (Eq. (18)), which can be in general nonlinear. The very good quality of the fits of the PVT data in the KABLJ model to Eqs. (13) and (14) with the functions $h(\rho)$ and $\gamma(\rho)$ given by Eqs. (19) and (20) has enabled us to very satisfactorily reproduce the dependences $B_T^{conf}(p)$ and $B_T(p)$ in the case of the KABLJ model, which have turned out to deviate gradually with increasing pressure from the high pressure extrapolations of their linear pressure dependences valid at low pressures.

We have suggested to employ Eqs. (21) and (22) respectively as the functions $h(\rho)$ and $\gamma(\rho)$ in Eqs. (13) and (14), because Eq. (4) with the function $h(\rho)$ given by Eq. (21) has been previously used[31] as the criterion for the generalized density scaling that has been met by real van der Waals liquids. Since there is no PVT data measured in the sufficiently wide pressure range, we have performed a preliminary indirect test of this assumption by using the high pressure dependence of the isothermal bulk modulus found from the ultrasonic autocorrelation measurements carried out[45] up to *ca.* $p$=1GPa. We have shown that Eq. (18) with the density function $\gamma(\rho)$ given by Eq. (22) yields a more reasonable ultra-high pressure prediction of the dependence $B_T(p)$ up to $p$=4GPa than that obtained from the typical quadratic pressure parameterization of $B_T(p)$.

In the case of real glass formers, the derived equations of state require further studying by using experimental volumetric data in the sufficiently wide pressure range, which are expected to be measured in the future. Then, if the structural relaxation or viscosity data are also accessible in the extremely wide pressure range, the functions $h(\rho)$ and $\gamma(\rho)$ suggested by Eqs. (22) and (23) can be even changed. Nevertheless, these EOS given by Eqs. (13) and (14) are proper and convenient templates for further ultra-high pressure investigations of volumetric properties of the materials approaching the glass transition, especially if their molecular dynamics obeys the generalized density scaling law given by Eq. (1).

**ACKNOWLEDGMENTS**

The authors gratefully acknowledge the financial support from the Polish National Science Centre within the program MAESTRO 2. K.K is deeply thankful for the stipend received within the project "DoktoRIS - the stipend program for the innovative Silesia", which is co-financed by the EU European Social Fund.